\documentclass[
superscriptaddress,twocolumn,
 amsmath,amssymb,
 aps,
]{revtex4-1}

\DeclareMathAlphabet{\mathpzc}{OT1}{pzc}{m}{it}
\newcommand{\G}{\epsilon}
\newcommand{\e}{\delta}

\usepackage{graphicx}
\usepackage{dcolumn}

\usepackage{bm}

\begin{document}

\preprint{APS/123-QED}

\title{Geometric localization in supported elastic struts}

\author{T.C.T. Michaels}
\thanks{These two authors contributed equally}
\affiliation{Paulson School of Engineering and Applied Sciences, Harvard University, Cambridge, MA 02138, USA}

\author{R. Kusters}
\thanks{These two authors contributed equally}
\affiliation{University Paris Descartes, Center for Research and Interdisciplinarity (CRI), 10 Rue Charles V, Paris}
\affiliation{Department of Applied Physics, Eindhoven University of Technology,\\ Eindhoven, The Netherlands}
\author{A.J. Dear}
\affiliation{Paulson School of Engineering and Applied Sciences, Harvard University, Cambridge, MA 02138, USA}
\author{C. Storm}
\affiliation{Department of Applied Physics, Eindhoven University of Technology,\\ Eindhoven, The Netherlands}
\affiliation{Institute for Complex Molecular Systems, Eindhoven University of Technology, Eindhoven, The Netherlands} 
\author{J.C. Weaver}
\affiliation{Paulson School of Engineering and Applied Sciences, Harvard University, Cambridge, MA 02138, USA}
\author{L. Mahadevan}
\thanks{Corresponding author: lmahadev@g.harvard.edu}
\affiliation{Paulson School of Engineering and Applied Sciences, Harvard University, Cambridge, MA 02138, USA}

\date{\today}

\begin{abstract}
Localized deformation patterns are a common motif in morphogenesis and are increasingly finding widespread applications in materials science, for instance as memory devices. Here we describe the emergence of spatially localized deformations in a minimal mechanical system by exploring the impact of growth and shear on the conformation of a semi-flexible filament connected to a pliable shearable substrate. 
We combine numerical simulations of a discrete rod model with theoretical analysis of the differential equations recovered in the continuum limit to quantify (in the form of scaling laws) how geometry, mechanics, and growth act together to give rise to such localized structures in this system. We find that spatially localized deformations along the filament emerge for intermediate shear modulus and increasing growth. 
Finally, we use experiments on a 3D printed multi-material model system to demonstrate that external control of the amount of shear and growth regulates the spatial extent of the localized strain texture.  

\end{abstract}

\maketitle


The deformation of spatially extended elastic structures, such as filaments, plates and shells is often elastically constrained by the surrounding medium. The simplest formulation of this elastic constraint is due to Winkler  \cite{Winkler1867} who proposed a local linear elastic model for the medium. Since then, there have been many variants that account for both nonlocality and nonlinearity, particularly in the context of localized deformations in such systems that take the form of creases, localized wrinkles etc. \cite{Cao2012, Jin2011,Takei2014,Hunt1998, Hunt2000, Pocivavsek2008, Yoo2008, Diamant2011, Brau2013,Auguste2014,Jin2015}. These localized patterns in spatially extended dynamical systems are of intrinsic interest due to their potential role as mediators of turbulence in fluids, optics and beyond  \cite{Knobloch2015,Pomeau_1986,Coullet_2000,Coullet_2003}. In the context of elasticity, a recent proposal suggests  the use of localized dimples as programmable elastic bits (e-bits) to store memory on a featureless elastic shell that requires minimal substrate infrastructure \cite{Chung_2018}, and raises the question of whether there are alternatives that take advantage of substrate elasticity to confine deformations and thus generalize these ideas to a much broader class of systems.

In this paper, we address this question  by studying the appearance of localized structures in a minimal model system of an extensible growing adherent filament that is attached to a rigid substrate by a set of stretchable and shearable springs. Using a combination of theory, simulations, and experiments, we show that this minimal system can give rise to robust spatial localization as a function of growth and shear.  More specifically, we derive explicit scaling relationships describing the extent and amplitude of localized deformations, as well as a phase diagram describing the range of mechanical parameters that admit localized structures. We then demonstrate these rules in practice using a 3D printed multi-material model system. Our results identify the physical parameters that control the extent of localization, suggesting a strategy for easy and robust programming mechanical deformations at multiple scales.

\begin{figure}[!ht]
\includegraphics[width=0.90\linewidth]{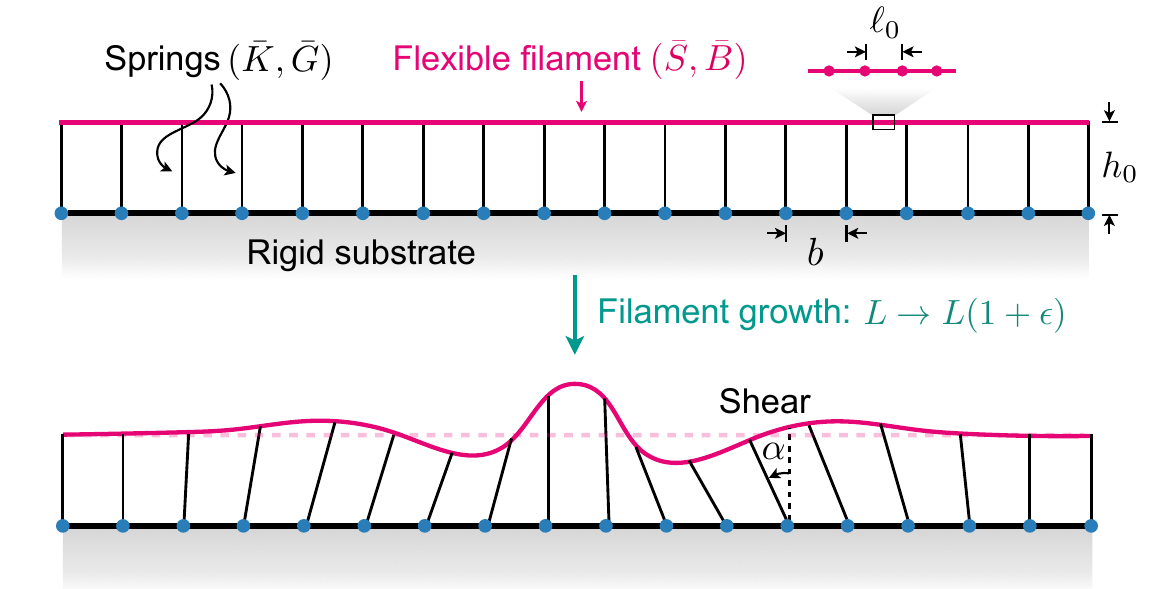}
\caption{Schematic representation of the discrete rod model of geometric localization. A flexible filament (with stretching stiffness $\bar{S}$ and bending stiffness $\bar{B}$) is attached to a rigid substrate through an array of springs that can extend vertically (with spring stiffness $\bar{K}$) and shear horizontally (with shear modulus $\bar{G}$). The the flexible filament undergoes uniform growth, such that its rest length $L$ increases to $L(1+\G) $, where $\G$ is the growth strain. Depending on shear modulus $\bar{G}$, filament growth $\G$ can result in the formation of localized deformation patterns. Understanding the conditions for the emergence of such localized deformations as a function of mechanical parameters is the subject of this paper.} \label{fig:1}
\end{figure}

\begin{figure*}[!ht]
\includegraphics[width=0.90\linewidth]{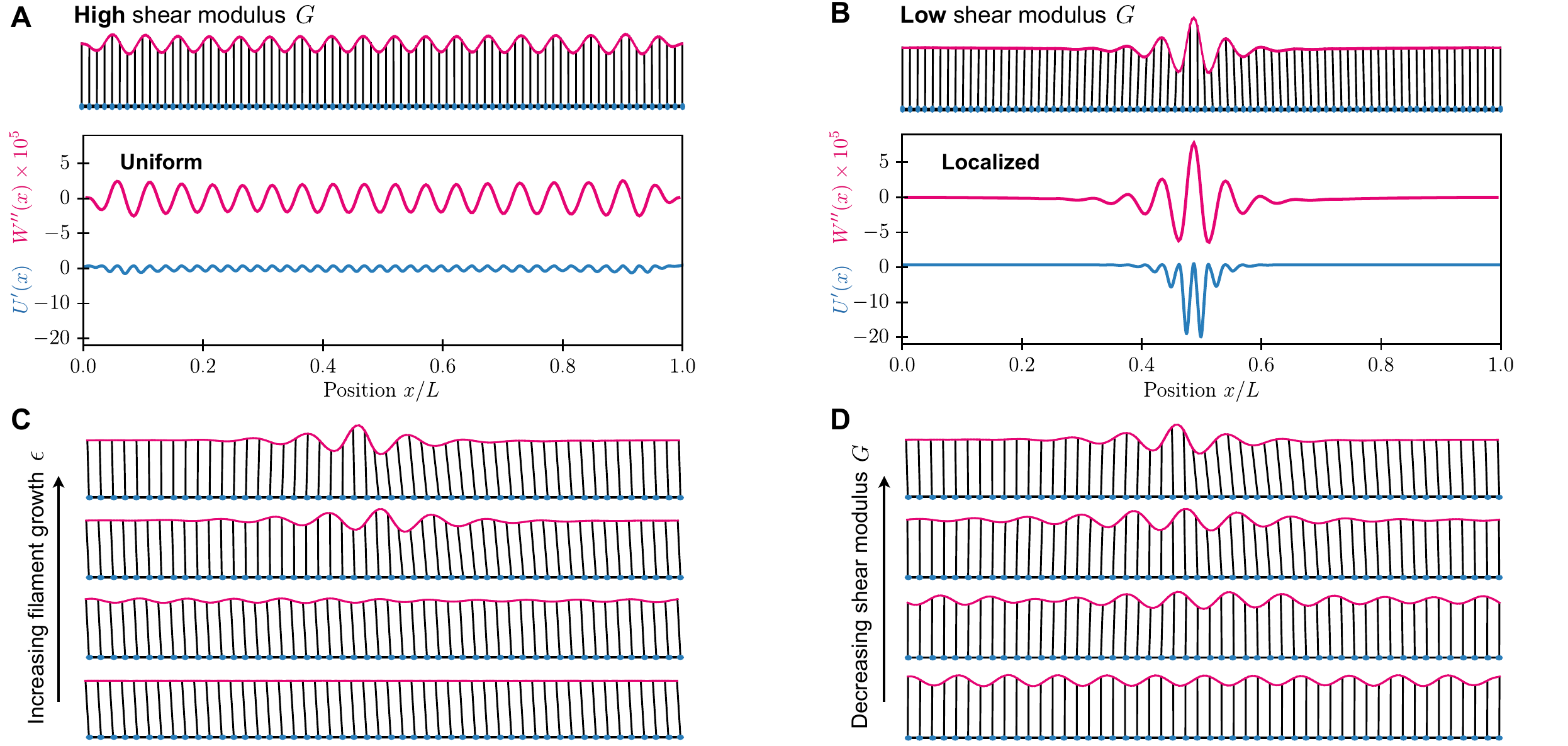}
\caption{Computer simulations of discrete rod model. (a-b) Depending on the shear modulus of the springs, filament growth results in uniform buckling for high shear modulus $\bar{G}=0.01$ (a), and localized deformations for low shear modulus $\bar{G}=0.0002$ (b). In both cases $N=100$, $n_s=316$, $b=1$, $h_0=1$, $\bar{K}=0.1$, $\bar{S}=100$, $\bar{B}=0.4$, and $\G=3.8\%$. (a) and (b) also show the curvature ($W''\times 10^5$, top purple line) and strain ($U'$, bottom blue line) fields along the filament. (c-d) Filament deformations for increasing growth strain $\epsilon$ (c), and decreasing shear modulus for the springs $\bar{G}$ (d). In (c) growth strain is $\G = 0.0004,$ $0.012$, $0.059$, and $0.075$ (bottom to top) at constant shear rate $\bar{G}=0.002$. In (d) the shear modulus is $\bar{G}=0.05,$ $0.01$, $0.005$, and $0.002$ (bottom to top) for $\G = 0.075$. In both (c) and (d) $N=50$.} \label{fig:2}
\end{figure*}

\section*{Theory}

\begin{small}
Our physical model of geometric localization is a flexible filament that can swell or grow, but is constrained by an elastic substrate to which it is attached. The filament can bend transversely and stretch longitudinally, while the substrate resists lateral shear and transverse stretch induced by the filament. We consider two limits of this problem: (i) a {\it{discrete}} version, where the filament is assumed to be made of a set of rigid links with soft connectors that resist bending, connected to a substrate via a set of discrete springs that resist changes in their natural length and orientation relative to the rigid rods (Fig.~\ref{fig:1}), and (ii) a {\it{continuum}} version, which maps onto the growth of a filament embedded in an elastic medium. In either case, we seek the solution with minimal total elastic energy for a filament with initial rest length $L$ which has grown/swollen to a total length of $L(1+\G)$, where $\G$ measures the relative growth (i.e.~strain) of the filament.

\subsection*{Discrete rod model}

The discrete rod model consists of a flexible filament made of $n_s$ discrete segments of rest length $\ell_0$ attached to a solid substrate through a set of $N$ Hookean springs that can extend vertically and shear horizontally (Fig.~\ref{fig:1}). The total energy of the discrete rod model  is
\begin{align}\label{Eq:model}
\mathcal{E}  &= \frac{\bar{S}}{2\ell_0} \sum_{n = 1}^{n_s}  \left( \ell_n - \ell_0 \right)^2 + \frac{\bar{B}}{\ell_0} \sum_{n=1}^{n_s-1} \left( 1 - \cos \theta_{n,n+1}\right) \\\nonumber
& \quad +  \frac{\bar{K}}{2b} \sum_{i = 1}^{N}  \left( h_i - h_0 \right)^2 +   \frac{\bar{G}}{2} \sum_{i = 1}^{N}   \alpha_i ^2,
\end{align}
where $\bar{S}$ denotes the filament stretching stiffness, $\bar{B}$ is the bending rigidity of the filament, $\bar{K}$ is the spring stiffness, and $\bar{G}$ is the shear modulus of the springs. Moreover, $\ell_n$ is the length of the $n$-th filament segment, $b = L/N$ is the spacing between the adherent springs, where $L=n_s \ell_0$ is the total length of the supporting filament.  In \eqref{Eq:model}, the first two terms correspond to the stretching and bending energies of the filament, while the last two terms correspond to the stretching and shear energies of the springs connecting the filament to the rigid substrate. The bending energy is consistent with the semi-flexible approximation valid for weak bending, where $\cos \theta_{i,i+1} = \hat{t}_i \cdot \hat{t}_{i+1}$, $i = 1,\dots, n_s$, where $\hat{t}_i$ is the unit tangent vector of the $i$-th filament segment. $\alpha_i$ denotes the angle that the $i$-th spring makes with the vertical, hence measuring the amount of shear; $h_i$ is the vertical extension of the $i$-th spring and $h_0$ is the rest length of the springs.

\begin{figure*}[!ht]
\centering
\includegraphics[width=0.90\linewidth]{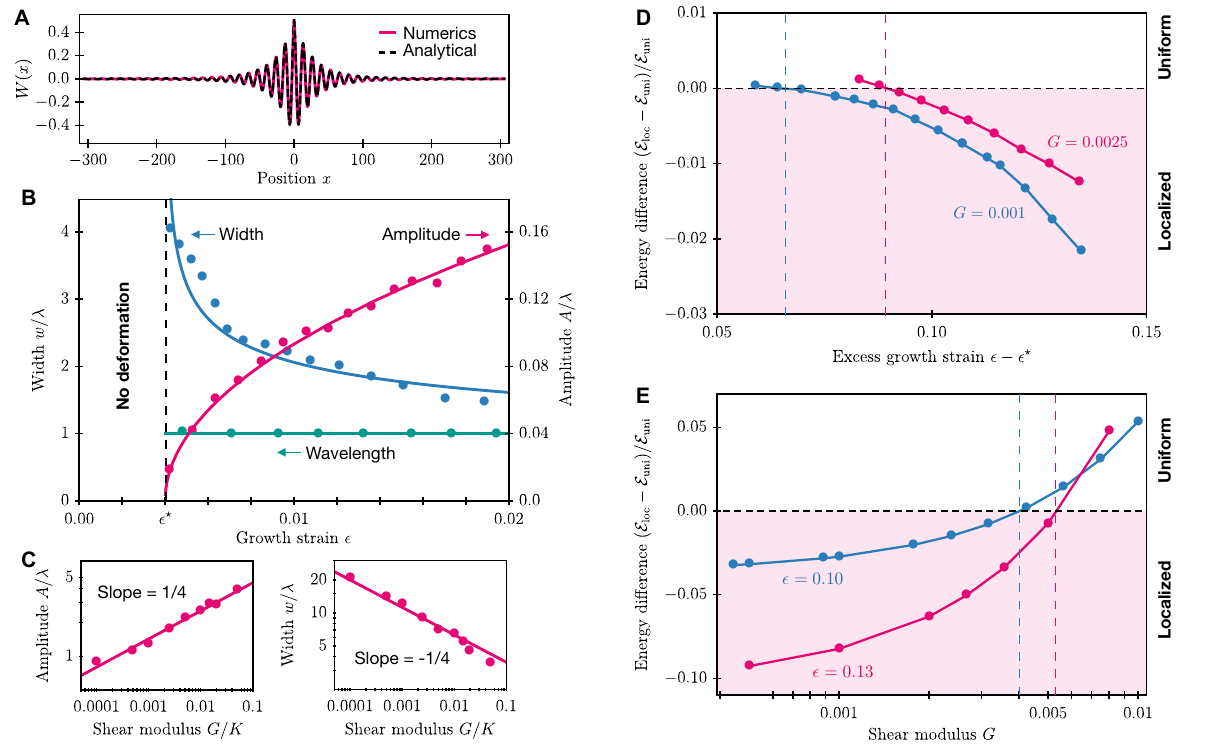}
\caption{Theoretical analysis of geometric localization. (a) Vertical displacement field $W(x)$ obtained by numerical integration of \eqref{Eq:EL3} (solid line) and comparison to the analytical perturbation solution, \eqref{W2} (black dotted line). Parameters: $G=5\times 10^{-5}$, $S=10$, $B=0.2$, $K=0.01$, $\G=9.5\times 10^{-3}$, $L=23\times \lambda$. 
(b) Bifurcation diagram for \eqref{Eq:EL3} as a function of growth strain $\G$. For $\G<\G^\star$ there is no deformation, while for $\G>\G^\star$ either uniform or localized deformations emerge. For $\G>\G^\star$, the amplitude of deformations increases in proportion to $\sqrt{\G-\G^\star}$, as predicted by \eqref{W2} (solid line). The figure also shows the width $w$ and wavelength $\lambda$ of localized deformations as a function of $\G$, with solid lines indicating the theoretical predictions. Data points are from numerical simulations of the discrete rod model for the following parameters: $N= 80$, $\bar{S} = 100$, $b = 1$, $n_s = 316$, $\bar{K} = 0.1$, $\bar{B}= 0.4$ and $\bar{G}=0.001$.  (c) Double logarithmic plots of amplitude $A$ and width $w$ of localized solutions as function of shear modulus $G$. The solid lines, which have slopes of 1/4, respectively, -1/4, are the theoretical predictions of \eqref{A} and \eqref{w}, $A \propto G^{1/4}$, $w \propto G^{-1/4}$. Simulation parameters: $S=10$, $B=1$, $K=1$, $\G=0.21$, $L=20\times \lambda$. (d-e) Uniform or localized deformations are selected depending on their relative energies, as described by $(\mathcal{E}_{\rm{loc}}-\mathcal{E}_{\rm{uni}})/\mathcal{E}_{\rm{uni}}$, where $\mathcal{E}_{\rm{loc}}$ is the energy of the localized state and $\mathcal{E}_{\rm{uni}}$ is the energy of the uniform state.
Panel (d) shows that, at constant $G$, there is a critical value for $\G$ above which the localized solution has lower energy compared to the uniform buckled solution. The critical $\G$ increases with increasing $G$.
Panel (e) shows that, at constant $\G$, there is a critical shear modulus $G$, below which the localized solution is lower in energy than the uniform deformation. The critical $G$ increases with increasing $\G$. Simulation parameters for (d-e) are: $N= 50$, $\bar{S} = 100$, $b = 1$, $n_s = 197$, $\bar{K} = 0.1$, $\bar{B}= 0.4$, $\bar{G}=0.001,0.0025$, and $\G=0.1,0.13$ as indicated on the graph.
\label{fig:3}}
\end{figure*}

The procedure we use to ``grow'' the filament is as follows: we initiate a configuration where the springs are in their rest configuration and vertical. We then increase the rest length of the filament segments $\ell_0$ by a small amount $\G \ell_0$, i.e.~$\ell_0 \to \ell_0(1+\G)$. Since thin rods are much easier to bend than to stretch, we focus our analysis on filaments with $\bar{S} \gg \bar{K}$; some extensibility is retained for increased numerical stability.
To minimize the total energy $\mathcal{E}$ of the system we either use Newton's
method, present in the FindMinimum routine, or a global energy search
present in the NMinimize routine. Both routines are implemented in Wolfram
Mathematica (see SI Sec.~S2).

\subsection*{Continuous model} 


To determine how localized filament deformations emerge during growth, we map the discrete rod model onto a set of coupled differential equations that describe the growth of the adhering filament in the continuum limit, valid for small deformations. To this end, we consider the elastic energy \eqref{Eq:model} in the limits $b,\ell_0\to 0$ and $N,n_s\to\infty$ with $N b = n_s \ell_0 =L$ constant. In this limit, the total elastic energy ${\cal E}$ of the system is written as a function continuous horizontal $U(x)$ and vertical $W(x)$ displacements, where $x$ is the arclength of the filament, as (see SI Sec.~S1):
\begin{align}
 {\cal E} & = \int_{L(1+\G)}  \left(\frac{S}{2} \left[U'(x)+ \frac{W'(x)^2}{2} - \G \right]^2\right.\nonumber\\
 & \qquad \left.+ \frac{ B}{2} W''(x)^2  + \frac{K}{2} W(x)^2 + \frac{G}{2} U(x)^2 \right)dx.
\label{Eq:L}
\end{align}
The elastic constants in the continuum model relate to those in the discrete model (denoted with bar) through $G= \bar{G} /(h_0^2 b)$, $K= \bar{K} /b^2$, $S=\bar{S} b/{\ell_0}$ and $B = \bar{B} /h_0$ (see SI Sec.~S1). The various energy terms in \eqref{Eq:L} have a straight-forward physical interpretation: the first term is the stretching energy of the filament and is proportional to the square of the elastic strain $\epsilon_{\rm{el}} = \epsilon_{\rm{tot}}-\G$, where $\epsilon_{\rm{tot}}$ is the total strain and $\G$ is the growth strain. The second term in \eqref{Eq:L}, proportional to the square of the local curvature $W''(x)$, describes the bending energy of the filament. The last two terms in \eqref{Eq:L} correspond to the energy contributions associated with the stretching and the shear of the substrate; these contributions are proportional to the squares of the vertical and horizontal displacement fields, respectively.

To find the system configurations of minimal total energy, we consider the Euler-Lagrange equations associated with \eqref{Eq:L}, $\delta {\cal E}/\delta W(x)=\delta {\cal E}/\delta U(x)=0$, which read (see SI Sec.~S1.6):
\begin{subequations}\label{Eq:EL3}
\begin{eqnarray}
BW''''(x)+ S\left[ \G - U'(x) - \frac{W'(x)^2}{2}\right] W''(x) & +   \\
 + KW(x) - G U(x) W'(x) &=& 0 \nonumber \\
S \left[ U''(x) + W'(x) W''(x) \right]-G U(x) &=& 0.
\end{eqnarray}
\end{subequations}
To compare with analytic results, we have used a shooting algorithm to  numerically solve \eqref{Eq:EL3} subject to clamped boundary conditions $ U(-L/2)=U(L/2)=0,\  W(-L/2)=W(L/2) = 0, \ W'(-L/2)=W(L/2) = 0$ (see SI Sec.~S2).
Note that these equations decouple and are fully integrable in two limits: (i) for $ G \to \infty$, and (ii) for $G=0$. Both situations lead to a uniform wrinkling solution with characteristic wavelength $\lambda = (B/K)^{1/4}$ (see SI Secs.~S3.2 and S3.3). 

\end{small}

\section*{Analysis}

\subsection*{Lowering shear modulus induces localization}

We have simulated the discrete rod model under different conditions of growth $\G$ and shear modulus $\bar{G}$ for clamped and free filament edges. For clamped edges, increasing $\G$ above a critical threshold $\G^\star$ and for large shear modulus $\bar{G}$, we observe a uniformly periodic deformation of the filament (Fig.~\ref{fig:2}(a)). This situation is consistent with the classical Winkler foundation \cite{Winkler1867}, corresponding to the $\bar{G}\to \infty$ limit where springs can extend only vertically. If we decrease $\bar{G}$, allowing for the adherent springs to hinge, we find that, upon uniform growth $\G$, the vertical displacement field along the filament is not uniformly periodic, but rather becomes localized (Fig.~\ref{fig:2}(b)). Since the filament itself is effectively inextensible ($\bar{S}/\bar{K}= 10^3$), this localized strain solution is accompanied by a shear field of the adherent springs (horizontal displacement). While the buckling wavelength remains unaltered, we find that localization is favored by increasing $\G$ (Fig.~\ref{fig:2}(c)) or decreasing $\bar{G}$ (Fig.~\ref{fig:2}(d)). Decreasing the shear modulus further leads to an increase of the localization width. Eventually, for vanishing shear modulus $\bar{G}=0$, we recover the uniformly deformed solution, when the width of the localized deformation approaches the system size $L$. 

Our simulations also point to an important role of the boundary conditions: releasing the edges results in the occurrence of localized structures that occur near the filament edge and upon further growth travel inward. In SI Sec.~S2 we discuss this aspect further.

\subsection*{Localized deformations as solutions to the continuous equations} 

To rationalize the emergence of spatial localized deformations in terms of specific combinations of the underlying physical parameters, we set out to solve the continuous equations, \eqref{Eq:EL3}, analytically in the limit of low shear modulus $G$ by employing weakly nonlinear analysis \cite{Drazin_2002}. We construct this solution on the half-axis $0\leq x\leq L/2$ by employing the mirror-symmetry of the vertical displacement around the midpoint $x=0$ (see SI Sec.~S3).

\begin{figure}[!ht]
\centering
\includegraphics[width=0.99\linewidth]{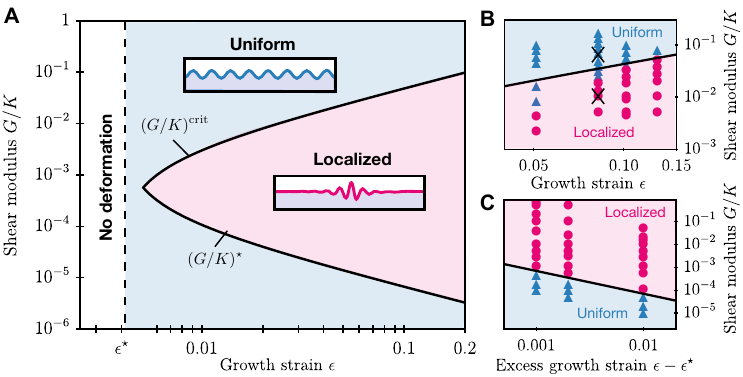}
\caption{Phase diagram as a function of shear modulus $G/K$ and growth strain $\epsilon$ separating regions where uniform or localized buckling deformations are minimum energy solutions. (a) In the purple region, bounded by the theoretical lines $(G/K)^{\rm{crit}}$ and $(G/K)^{\star}$, localized deformations exist and have lower energy than uniform solutions. Diagram calculated for the same parameters as in Fig.~\ref{fig:3}(d). (b) Phase diagram in high $G$ regime obtained from numerical simulations of discrete rod model for the parameters of Fig.~\ref{fig:3}(d) and comparison with theoretical prediction (solid line). Circles indicate localized deformations, triangles indicate uniform deformations. At the crosses we detected bistability. (c) Phase diagram in low $G$ regime obtained from numerical simulations of continuum model for the parameters of Fig.~S5(b) and comparison to the theoretical prediction \eqref{star} (solid line, which has slope -1).\label{fig:3b}}
\end{figure}

\subsubsection*{Linear theory and bifurcation analysis} 

As a first step, we study the linearized equations $
BW''''(x) +  S \G W''(x)+KW(x)  =0$, $S U''(x) =G U(x)$
subject to the boundary conditions associated with clamped edges (see SI Sec.~S3.1). For $\G \leq \G^\star$, where
\begin{equation}
\G^\star = \frac{2\sqrt{BK}}{S},
\end{equation}
we find that the only solution consistent with the boundary conditions is $W(x)=U(x)\equiv 0$, i.e.~no deformation. However, for $\G \geq \G^\star$, we find a non-trivial, uniformly periodic wrinkling solution for the vertical displacement; the wavelength of wrinkling is $\lambda =(B/K)^{1/4}$, consistent with  previous reports \cite{Landau, Cerda_2003}. Thus, analysis of the linearized equations implies a bifurcation point when $\G=\G^\star$ (Fig.~\ref{fig:3}(b)). The physical interpretation of $\G^\star$ is the maximal amount of strain that can be accommodated by filament stretching (compression) without causing any bending and spring deformation. Indeed, $\G^\star$ is inversely proportional to $S$; decreasing $S$ leads to a higher strain threshold $\G^\star$. Above the critical swelling/growth strain $\G^\star$,  the excess strain $\G-\G^*$ is accommodated by filament buckling.

\subsubsection*{Weakly nonlinear theory} 

The linear theory implies that localized deformations exist for \eqref{Eq:EL3} only for $\G>\G^\star$; we thus consider the non-linear terms in \eqref{Eq:EL3} as a perturbation of the linearized equations in excess strain $\G-\G^\star$. To do so, we rescale variables as $\bar{x}=x/\lambda$,  $W =\sqrt{\G-\G^\star}\, \lambda \, \bar{W}$ and $U(x) =(\G-\G^\star)\, \lambda \, \bar{U}$, solve the resulting equation for $U(x)$ exploting $(G/S)\sqrt{B/K}\ll 1$ and insert this solution back into \eqref{Eq:EL3}. This procedure (detailed in SI Sec.~S3.5) yields at leading order:
\begin{align}\label{eqs_3}
\bar{W}''''(\bar{x}) + & 2\bar{W}''(\bar{x})+\bar{W}(\bar{x})  - \e\frac{(\bar{x}-\bar{L}/2)^2}{2}\bar{W}''(x) =0,
\end{align}
where the relevant perturbation parameter of the problem emerges as
$\e = (G/K)(\G-\G^\star)\ll 1$. In \eqref{eqs_3} we recognize the linear part, which yields uniform buckling with wavelength $\lambda$; as expected, we obtain a new term, proportional to $\delta$, which comes from the excess strain $\G-\G^\star$ and varies slowly on distances of scale $\lambda$. This problem is analogous to the WKB approximation to the Schr\"odinger equation with a slowly varying potential.
The form of \eqref{eqs_3} thus suggests the following envelope-type Ansatz for the vertical displacement
\begin{equation}\label{ansatz}
\bar{W}(x) = A(\e^{1/4}\bar{x})e^{i \bar{x}} +\textrm{c.c.}
\end{equation}
where $A$ is a slow-varying amplitude (envelope) function that depends on the slow variable $X=\e^{1/4}\bar{x}$ and c.c.~stands for complex conjugate. Inserting this Ansatz into \eqref{eqs_3} and collecting terms at various orders in $\e^{1/4}$, we arrive at the following amplitude equation describing the long-scale behavior of the solution (see SI Sec.~S3.6):
\begin{equation}
\frac{\partial^2 A(X)}{\partial X^2}= \frac{(X-X_{e})^2}{8}A(X),
\end{equation}
where $X_{e} = \e^{1/4} L/\lambda$. The amplitude equation is a particular case of the Weber differential equation
$y''(z) + \left(\nu+1/2-z^2/4\right)y(z)=0$ with $\nu=-1/2$, whose solution is expressed in terms of the parabolic cylinder function $y(z) = D_{-1/2}(z) $ \cite{Abramowitz}. Hence, the final perturbation solution for the vertical displacement is given by:
\begin{align}\label{W2}
W(x) & \propto  \sqrt{\G-\G^\star}\lambda\, \cos\left(\frac{x}{\lambda}\right)\, \times \\\nonumber 
& \quad D_{-1/2}\left[(\G-\G^\star)^{1/4}\left(\frac{G}{2B}\right)^{1/4}\left(x-\frac{L}{2}\right)\right].
\end{align}
The solution has the form of the uniform wrinkled solution $\cos(x/\lambda)$ modulated by an envelope function $A(X)$, which depends on the slow variable $X=(\G-\G^\star)^{1/4}(G/B)^{1/4}x$. Note that, independently of growth strain $\G$ or shear modulus $G$, the solution always selects the wavelength $\lambda$. The accuracy of \eqref{W2} against numerical integration of \eqref{Eq:EL3} is shown in Fig.~\ref{fig:3}(a) and in SI Figs.~S7-S9.

 \begin{figure*}[!ht]
\centering
\includegraphics[width=0.99\linewidth]{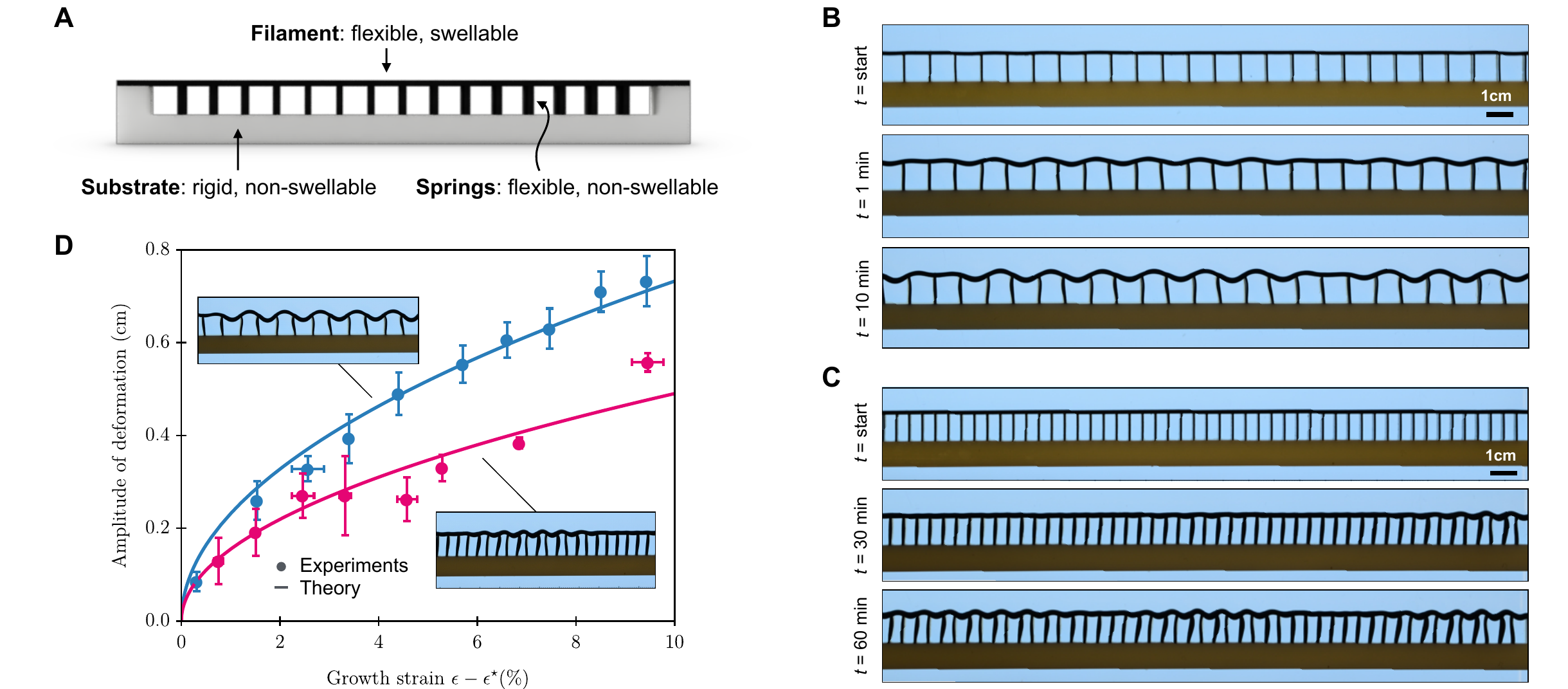}
\caption{Experimental realization of geometric localization using 3D printing. (a) Schematic representation of the experimental realization of the system using a multi-material 3D printed plastic model for the growth of a filament attached to a shearable substrate. The ends of the flexible filament are clamped. (b-c) Images of swelling 3D-printed samples placed in isopropyl alcohol showing the development of localized deformations over time: the number of vertical springs is $n_s=28$ (b), and $n_s=56$ (c). The different spacing between ``springs'' allows us to control the mechanical parameters $G,K,S$, hence the form of localized deformations. Scale bar is 1 cm. (d) Measured amplitude of deformations for the samples in (b) and (c) as a function of growth strain $\G-\G^\star$ is compared to theory, \eqref{A} (solid line). \label{fig:4}}
\end{figure*}

\subsubsection*{Scaling laws for localized solution}

With a perturbative solution to the continuous equations \eqref{Eq:EL3} at hand, we can now predict, from first principles, the scaling behavior of a number of key observables, such as the width $w$, the wavelength $\lambda$, and the amplitude $A$ of the localized deformations. The resulting scaling relationships provide a useful means for designing localized deformations across different scales. In particular, from \eqref{W2}, the width $w$ of the localization scales as:
\begin{equation}\label{w}
\frac{w}{\lambda} \propto \delta^{-1/4} = (\G-\G^\star)^{-1/4}\left(\frac{G}{K}\right)^{-1/4}.
\end{equation}
Hence, increasing growth strain $\G$ beyond the bifurcation point $\G^\star$, or increasing $G$ favors localization (Fig.~\ref{fig:3}(b,c)). The wavelength of the localized deformation $\lambda = \left(B/K\right)^{1/4}$ is independent of growth strain $\G$ and shear modulus $G$ (Fig.~\ref{fig:3}(b)).
Finally, the amplitude $A$ of the localization scales as
\begin{equation}\label{A}
\frac{A}{\lambda} \propto (\G-\G^\star)^{1/2}\left(\frac{G}{K}\right)^{1/4} \left(\frac{L}{\lambda}\right)^{1/2}.
\end{equation}
We have tested these scaling predictions using numerical realizations of the continuous equations obtained with a shooting algorithm. As shown in Fig.~\ref{fig:3}(b,c) and Fig.~S10, the scaling relationships of \eqref{w} and \eqref{A} with $\G$ and $G$ are verified by numerical analysis, thus confirming the power of our theory to capture the key mechanical parameters that determine the onset and the characteristics of localized solutions.

\subsection*{Energetics of localized solution}

We also studied the energetics of the single localized and uniform buckled solutions (the energetics of higher-order solutions is discussed in SI Sec.~S2; also see SI Sec.~S2 for details on energy calculation procedure). In Fig.~\ref{fig:3}(d), we plot the energy difference between localized and uniform buckling deformations, $(\mathcal{E}_{\rm{loc}}-\mathcal{E}_{\rm{uni}})/\mathcal{E}_{\rm{uni}}$, as a function of excess strain $\G-\G^\star$ at fixed values of shear modulus $G$. We find that there is a critical strain $\G$ above which the localized solution has lower energy than the uniform solution; this critical strain increases with increasing $G$ or decreasing $K$. In fact, the energy difference between the localized and the uniform states is set by an interplay between shear energy and extensional energy of the substrate springs; localizing the strain of the filament increases the amount of shear, while decreasing the extensional energy of the substrate springs (see SI Sec.~S3.7).

Similarly we can also pick a solution for a given $\G$ and study how varying $G$ impacts its total energy. Since we do not perturb the solution it will remain trapped in the local minimum. We find that below a critical $G$, the energy difference between localized and uniform states becomes negative, i.e.~the localized state has a lower energy compared to the uniform solution (Fig.~\ref{fig:3}(e)). The critical $G$ increases for increasing $\G$, hence expanding the window of values for shear modulus for which we can expect to find localized states.

\subsection*{Phase diagram for localized deformations}

In summary, our analysis implies that localized deformations are stable solutions for intermediate values of shear modulus $G$ (Fig.~\ref{fig:3b}). By analyzing the energy of uniform and localized solutions, we found that decreasing $G$ from the $G\to \infty$ limit below a critical value $G^{\rm{crit}}$ causes localized deformations to have a lower total energy than the homogeneous buckled solution. This critical value is given by (see SI Sec.~S3.7) \begin{equation}
(G/K)^{\rm{crit}} \propto \G-\G^\star;\end{equation} it increases with increasing $\G$. Moreover, our perturbative solution, \eqref{W2}, valid for low $G$, shows that, in this limit, decreasing $G$ causes the width $w$ of localization to increase; eventually, localization disappears when $w$ approaches system size $L$; using \eqref{w}, we find that this occurs below \begin{equation}\label{star}  (G/K)^{\star} \propto (\G-\G^\star)^{-1}(\lambda/L)^4.\end{equation}  These two effects give rise to the phase diagram in Fig.~\ref{fig:3b}(a), where localized deformations emerge for intermediate values of $G$. We have verified these relationships using computer simulations both in the high $G$ (Fig.~\ref{fig:3b}(b)) and low $G$ regimes (Fig.~\ref{fig:3b}(c)). 

We note that our system displays bistability. If we prepare the homogeneous state at $G = 0.001$, which is well below the value at which it is the global minimum (Fig.~\ref{fig:3b}(b)), we find that random perturbations of the filament, up to the buckling amplitude, relax back to the homogeneous state. We have similarly verified the stability of the localized state at $G = 0.005$ and found that this state relaxes back for amplitudes of perturbations up to the buckling amplitude. At higher values of $G$ we recover homogeneous buckling or less pronounced localized buckling.

\section*{Experiments}

Using a multi-material 3D printer (Stratasys Connex500), we constructed a simple experimental realization of a growing filament bound to a shearable substrate to asses the occurrence of localized deformations (Fig.~\ref{fig:4}(a)). The 3D printed samples consisted of three parts: (i) a rigid, non-swellable substrate, (ii) a flexible and swellable ``filament'', and (iii) a series of flexible and non-swellable ``springs'' connecting the filament to the substrate \cite{Guiducci_2015}. During the fabrication process, a photosensitive liquid precursor (the 3D printer ink) is deposited in a voxel-by-voxel fashion. Several precursors are used to print multiple materials with different properties and the resulting modulus can be tuned by varying the concentration of photo-initiator. A UV light cross-links the liquid precursors in a layer-by-layer fashion and this process is repeated until the full 3D model is built. Depending on the liquid precursor composition and the degree of cross-linking, a broad range of mechanical properties can be achieved from stiff thermoplastic-like to soft rubber-like materials. The degree of cross-linking also directly influences the swelling capacity of the polymer, and for the experiments described here, the component formulations were guided by previous formulations developed by Guiducci, et al.~\cite{Guiducci_2015}. Growth of the filament is induced through swelling by submerging the sample in isopropyl alcohol, and the sample deformations were filmed over 1 hour with a digital camera. The structures were elevated off the bottom of the container to ensure that no flexible component was in direct contact with the walls, which would cause friction effects that could limit the motion of the filament or the springs. The swelling experiments were conducted on two different geometries with the same filament length $L$, but consisting of $n_s=28$ and $n_s=56$ vertical springs (Figs.~\ref{fig:4}(b,c)), i.e.~the spacing $b$ between springs is changed by a factor of 2 between the two experimental realizations. In both cases, $b$ was sufficiently large to avoid differences in solvent diffusion between the two samples. Since the material properties as well as thickness of all components were identical for both realizations, changing $b$ allows us to control the stretching modulus of the filament, $S=\bar{S}b$, as well as the stretching and shear moduli of the springs, $K=\bar{K}/b^2$ and $G=\bar{G}/(h_0^2b)$, while keeping the other parameters unchanged. According to our theory, the width, wavelength and amplitude of localized deformations scale as $w\propto G^{-1/4}$, $\lambda \propto K^{-1/4}$ and $A\propto (G/K)^{1/4}$; hence, we predict that reducing $b$ results in localized deformations with overall smaller width, wavelength and amplitude. These predictions agree with the experimental observations (Figs.~\ref{fig:4}(b,c)). Moreover, the critical strain for the onset of deformations scales as $\G^\star = \sqrt{BK}/S$; hence decreasing the spacing $b$ between springs is expected to increase $\G^\star$. This prediction is consistent with the experimental observation that sample (c) requires a longer time for the onset of localized deformations compared to (b). For both geometries, we measured the amplitude of deformations as a function of excess growth strain $\G-\G^\star$ and verified the square-root scaling of the bifurcation predicted by \eqref{A} (Fig.~\ref{fig:4}(d)). 

\section*{Discussion}
 
We have described and studied a minimal realization of spatially localized deformations in a growing filament connected to a shearable substrate using a discrete rod model, a continuum theory, and implemented the results using experiments using a multi-material 3d printing framework. Our main results take the form of scaling relationships for key parameters relating to localized deformations, such as amplitude, wavelength, and width, and a phase diagram for uniform and localized deformations. These can be particularly useful as guiding principles for designing controlled deformations with potential applications to mechanical memories. Indeed, our work shows that the ability to achieve localized deformations in filaments bound to shearable substrates depends on specific combinations of the mechanical parameters and the size of the system. Upon appropriate rescaling using the scaling relationships derived in this study, localized deformations could be appropriately designed both at the macro- and micro-scales.

Our work complements previous extensive mathematical literature on localized deformations in a variety of non-equilibrium settings \cite{Pomeau_1986,Coullet_2000,Coullet_2003,Swift1977,Cross1993,diBernardo2008,Devaney1977,Lee1993}, by providing a simple physical realization of such localized deformations in equilibrium mechanical systems in a one-dimensional setting, complementing earlier work by us in two-dimensional settings \cite{Chung_2018}.  Our work also provides an example of how to employ multi-material 3D printing as a reproducible, rapid and easily realizable method for studying complex problems related to mechanically constrained growth at multiple scales. Due to the ability to control the deposition of material with very high resolution to create complex 3D structures, this approach could thus provide a practical research platform for investigating mechanical feedback on growth under different conditions, e.g.~in the presence of gradients in material properties. By incorporating extensions to 2D systems, our approach could move us one step closer to understanding the two way-feedback between mechanics and growth/swelling kinetics, which is a defining feature in the growth of spatially extended structures in both materials science and biology.

\section*{Acknowledgments} 
We acknowledge financial support from the Swiss National Science foundation (TCTM), the Netherlands Organization for Scientific Research (NWO-FOM) within the program “Barriers in the Brain: the Molecular Physics of Learning and Memory” (No.FOM-E1012M, RK and CS), and the NSF BioMatter DMR 33985 (LM) and MRSEC  DMR 14–20570 (LM). We thank Nicholas W.~Bartlett for technical support during the swelling experiments.

\end{document}